# Constraining the nature of galaxy haloes with gravitational mesolensing of QSOs by halo substructure objects

© Yu. L. Bukhmastova [1], Yu. V. Baryshev[2]

[1] Saint-Petersburg State University, Russia, Email: bukh_julia@mail.ru
[2] Saint-Petersburg State University, Russia, Email: yuba@astro.spbu.ru

**Abstract:** Gravitational lensing of background compact objects like active galactic nuclei and quasars, by intermediate mass lenses such as globular clusters and and dark matter clumps with masses $10^5 - 10^8$ $M_\odot$, is considered. It is shown that observational study of the galaxy-quasar's associations is a powerful direct observational test of the nature of massive galaxy haloes. Optical interferometric observations with VLTI and Keck instruments are able to constrain masses and number of substructure halo objects. Evidence of gravitational lensing by globular clusters in haloes of spiral and elliptical galaxies is presented.

## 1. Mesolensing as a probe of the halo substructure

The theory of gravitational lensing was well described by Bliokh and Minakov [8], Zakharov [35], and Schneider et al. [26]; many papers are devoted to quasar–galaxy associations (see, e.g., Thomas et al. [31]; Zhu Zong-Hong et al.[36]; Benitez et al. [5]; Menard and Bartelmann [22]; Scranton et al. [27]; for the possibility of strong gravitational magnification of quasars, see Yonehara et al. [34]).The idea that halo objects could act as lenses was used by Barnothy [1], with globular clusters, Canizares [13], with dark condensed objects, Baryshev et al.[4] , with globulars and other intermediate (meso) mass objects, to explain properties of galaxy-quasar pairs. Then Baryshev & Ezova [3] calculated probabilities of strong lensing of compact background objects by lenses with King-type mass distribution which populate galaxy haloes. The lenses were taken to be globular clusters, dwarf galaxies and clusters of dark matter with masses of $10^3-10^9 M_\odot$. They pointed out that a consequence of the mesolensing would be splitting or spreading of the star-like image to 10 – 100 milliarcsecond size. Recently, Yonehara et al. [34] arrived at similar conclusions, now as a response to the CDM predictions by Klypin et al. [20] and Moore et al. [23] on large number of substructures within galaxy haloes. We present hereafter some simple relevant calculations. The typical expected angular separation between multiple quasar images is: $\Theta = (4GM/c^2)^{1/2} D_{eff}^{-1/2}$, where $D_{eff} = (D_{ol}D_{os}/D_{ls})$ is expressed in terms of the angular size distances between the observer, the lens, and the background source (quasar). Normalized to a mass of $10^8 M_\odot$, with $D_{eff}$ in Mpc, we obtain for $\Theta$ (in mas): $\Theta = 100(M/10^8 M_\odot)^{1/2}(D_{eff}/100Mpc)^{-1/2}$. Thus we see that lumps of $10^8 M_\odot$ can produce images of $0.1''$, when the effective distance of the lens galaxy is 100 Mpc. One may make a rough estimate of the probability of mesolensing for a quasar image located near the line-of-sight towards a foreground galaxy halo: $P = N[(4GM/c^2)/R^2_{halo}]D_{eff}(D_{ls}/D_{os})^2 Bias(V)$, where $Bias(V)$ represents the magnification bias correction as a function of the magnitue $V$ of the background source [29]. If we simply assume that the halo is made out of $N_i$ lumps of mass $M_i$, then this may be conveniently expressed in terms of the halo mass $M_{ha}$ and size $R_{ha}$: $P = 0.1(M_{ha}/10^{12}M_\odot)/(R_{ha}/0.1Mpc)^2(D_{eff}/100Mpc)(D_{ls}/D_{os})2Bias(V)/Bias(V=15)$

## 2. Cold dark matter halo structure

In theoretical cosmology, cold dark matter (CDM) plays the role of primordial substance, which via gravitational clustering hierarchically gathers into lumps, finally forming the bodies of galaxies on which luminous baryonic matter is concentrated. Cosmological N-body simulations of clustering are able to produce large-scale structures rather similar to those observed (groups, clusters, superclusters), which is justifiably regarded as a triumph for this original theory. Until recently, it was not possible to achieve the computational resolution to follow what happens to the CDM within galaxy haloes. Extremely interesting results for these small scales were obtained in 1999 by Klypin et al. [20] and Moore et al. [23]. In particular, Moore et al. note that high-resolution simulations lead to the picture where "galaxies are scaled versions of galaxy clusters", so that the massive haloes contain large numbers of CDM lumps as sate lites of the galaxy. These lumps may have masses in the range $10^6 - 10^9$ $M_\odot$ and galaxies with a Milky Way halo mass could have 5000 dark satellites with masses larger than $10^8$ $M_\odot$. These theoretical entities are a challenge in the standard picture of CDM structure formation and attempts to observationally prove or disprove their existence form an important test of the CDM scenario.

## 3. Description of the proposed programme

Gravitational mesolensing of background quasars by intervening galaxy halo objects may be used as a tool to test the above fundamental prediction of the CDM model. We suggest to observe with the VLT (NAOS/CONICA) a sample of bright quasars whose lines-of-sight pass through the haloes of foreground galaxies (projected distance less than 100 kpc). Our first sample of quasar-galaxy pairs are so selected that the expected splitting of the image is about 100 mas or larger for gravitating lumps of $10^7 M_o$. By selecting bright quasars ($V < 16$ mag) we automatically increase the probability of including lensed images because of the magnification bias [29], which may be a few magnitudes. In fact, in the Hubble diagram, these quasars are brighter than the AI quasars as defined by Teerikorpi [30], which should be unlensed, optically quiet most luminous quasars.

We first observe these objects where possible lenses should be the easiest to detect, so that in case of positive detection we can increase the sample for the next phase. Also, the sample is so selected that in case of all negative detections, we may already start constraining relevant quantities, such as the masses of halo lumps. It is first enough to measure the angular size of the elongated image without detailed structure information. Because of the necessity to restrict to rather nearby galaxies in order to have a large enough angular image size, $D_{eff}$ at most a few hundreds Mpc, the amplified lensing probability becomes of the order of 0.1. If lensing and amplification significantly contribute to the well-known excess of high-z quasars close to low-z galaxies [3], then it is clear that in observed samples of galaxy-quasar pairs mesolensing must be common. In future, with a better angular resolution, one may select pairs with larger lensing probabilities.

We have taken pairs with $V < 16$ for quasars, the projected quasar-galaxy distance < 100 kpc and the image size for substructure mass of $10^7 M_o$ about 100 mas.

**List of targets proposed in this programme**

| Target/Field | α(J2000) | δ(J2000) | $m_v$ | Diam. Q (arcsec) |
|---|---|---|---|---|
| PKS 1004+13 | 10 7 26.2 | 12 48 56 | 15.2 | 0.141 |
| PG 1008+133 | 10 11 10.8 | 13 04 12 | 16.3 | 0.141 |
| HE1104-1805 | 11 06 33.6 | -18 21 25 | 16.2 | 0.035 |
| PG 1216+069 | 12 19 20.9 | 6 38 38 | 15.6 | 0.100 |
| PG 1254+047 | 12 56 59.9 | 4 27 34 | 16.2 | 0.078 |
| PKS 1355-41 | 13 59 0.2 | -41 52 53 | 15.9 | 0.097 |

## 4. Quasars lensed by globular clusters of spiral and elliptical galaxies

Quasars are observed in pairs with galaxies of various types. If distant quasars and nearer galaxies are associated via halo globular clusters, then the distribution of quasars around elliptical galaxies must follow in a certain way the distribution of globular clusters around these galaxies. Does the distribution of quasars around elliptical and spiral galaxies have peculiarities of its own [12]?

To select quasar–galaxy pairs, we use the available catalog of 8382 pairs [9] compiled from the LEDA database [24] and the catalog of quasars [32]. We also select new pairs from the fourth version of the SDSS catalog of quasars and galaxies (www.sdss.org). The selection criteria are the following.
1. We select quasars and galaxies with available data on their spatial coordinates α, δ, and z. The apparent magnitudes are also known for the quasars.
2. Each quasar in a pair must be farther than the galaxy, i.e., $z_{qso} > z_{gal}$.
3. The quasars must be projected onto the galaxy halos. The halo size does not exceed 150 kpc.
4. We consider nearby quasars with $z < 0.3$.
5. We select pairs with $z_{gal}/z_{qso} > 0.9$.

We will call the pairs with $a > 0.9$ close quasar–galaxy pairs, because the quasar and the galaxy in such a pair are close to one another not only in angular separation, but also in redshift. Criterion 4 also reduces the sample by removing the distant quasars that can be paired with distant galaxies, which are undetectable in the surveys under consideration.

Bukhmastova [9] selected 8382 pairs based on criteria 1–3. Now we selected 64 close quasar–galaxy pairs with the additional constraints 4 and 5. We will call the sample of pairs based on the LEDA database and the catalog of quasars [32] sample 1 [12].

Let us select pairs based on the fourth version the Sloan Digital Sky Survey (www.sdss.org). 5224 quasars with $z < 0.3$ and 321 516 galaxies are involved in the selection. We found the 64 pairs and call the sample 2 [12].

## 5. Analysis of the selected pairs

Let us analyze the derived samples of close quasar–galaxy pairs. Let us determine the linear distances from the galaxy centers at which the quasars from pairs are projected. Figures 1 and 2 show the distributions of quasars from samples 1 and 2, respectively. The distance from the galaxy center to the quasar projection in kpc is along the horizontal axis and the number of quasars is along the vertical axis.

Sample 1 of close quasar–galaxy pairs is represented mostly by spiral galaxies. Our main assumption is that the quasars of sample 1 are associated with spiral galaxies via globular clusters in the halos of these galaxies. Thus, Fig.1 leads us to conclude that 80% of the globular clusters in spiral galaxies are located in their halos at distances up to 40 kpc. Note that the clustered halo objects are predominantly within 10 kpc; further out, the number of such objects decreases sharply. Below, we will establish whether this is actually the case.

To establish the types of galaxies represented in sample 2, let us turn to the paper by Fukugita et al. [16], who classified the galaxies according to their color indices. We conclude that about 70% of the galaxies in the pairs of sample 2 are elliptical ones. In the section on globular clusters in the elliptical galaxies A, A1644, A2124, A2147, A2151, A2152 of the Abell cluster and in the galaxies IC 4051, M49, and M87, we will establish whether this means that the data in Fig.2 reflect the spatial distribution of globular clusters in elliptical galaxies.

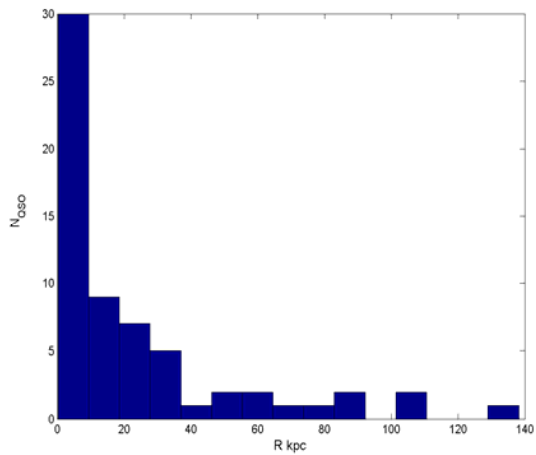
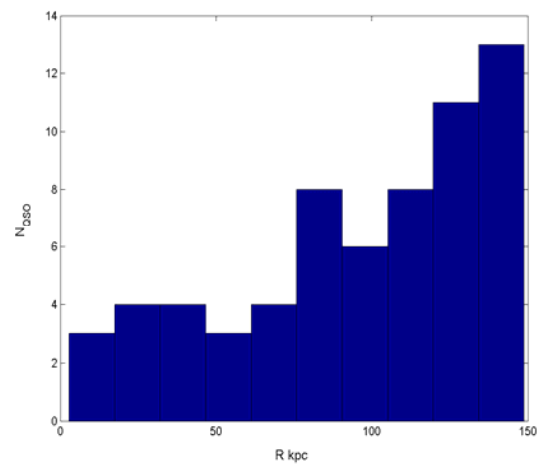

**Fig. 1.** Distribution of 64 quasars from sample 1 relative to the centers of the galaxies onto the halos of which they are projected.

**Fig. 2.** Same as Fig. 1 for sample 2.

## 6. Globular clusters in spiral galaxies and quasar-galaxy associatios

To establish the peculiarities of the distribution of globular clusters in the halos of spiral galaxies, we use data on the locations of 150 globular clusters in the Milky Way halo [18]. Analysis of these data is presented in Fig.3. It follows from this analysis that more than 80% of the globular clusters are actually located at distances of no larger than 40 kpc. To determine the density profile of globular clusters in the halos of spiral galaxies in more detail, let us turn to the data on 1164 candidates for globular clusters in the Andromeda Galaxy [17]. Figure 4 shows the distribution of these globular clusters in projection onto the galactic plane. The distances from the galaxy center (point 0.0) to the globular clusters are along the axes. We counted the halo globular clusters in rings of a fixed radius. The upper straight line in Fig.5 represents the density profile of globular clusters in the Andromeda Galaxy. We see that the radial density of globular clusters in the halos of spiral galaxies is well described by a power law in the form $n_{GC}(r_p) = A_0 * r_p^{\alpha}$, where $\alpha=-2.1 \pm 0.3$ and $\alpha=-2.5\pm 0.3$ for the Andromeda and the Milky Way, respectively; $n_{GC}$ is the number of globular clusters per unit area of the halo ring; and $r_p = R$ is the galactocentric distance of the globular cluster in projection onto the galactic plane.

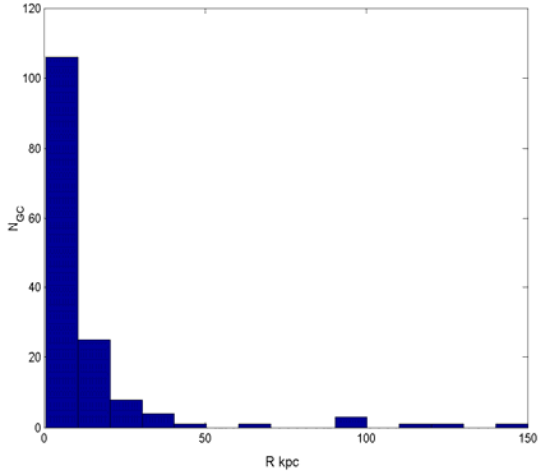

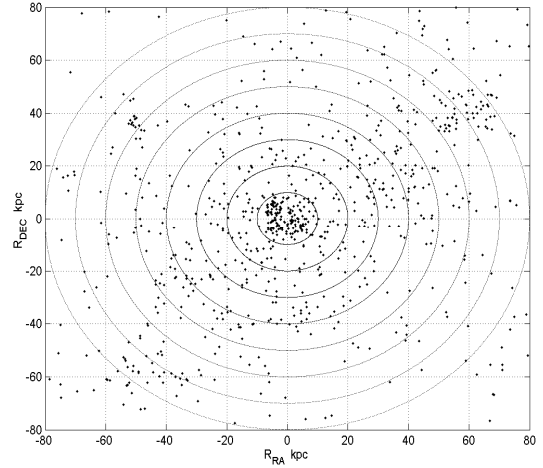

**Fig. 3.** Distribution of 150 Milky Way globular clusters relative to the Galactic center

**Fig. 4.** Distribution of globular clusters of the Andromeda Galaxy in the halo plane. The equatorial coordinates recalculated to galactocentric distances are along the axes.

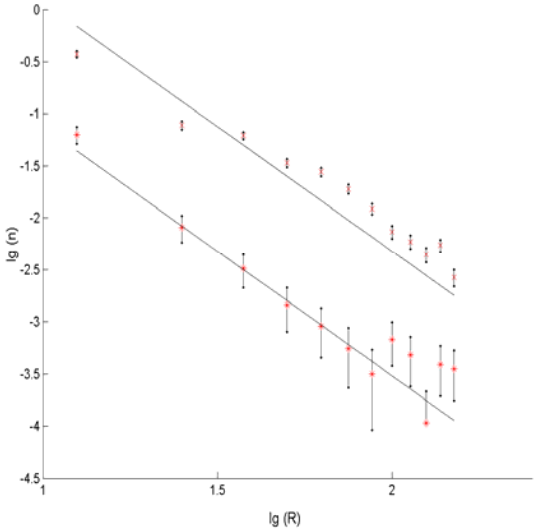

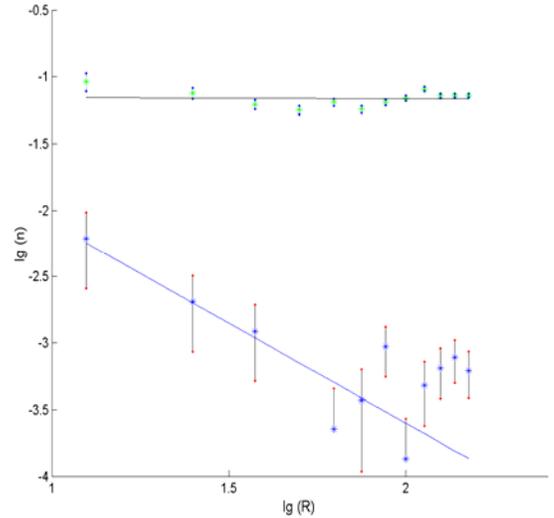

**Fig. 5.** Density profiles of globular clusters in the Andromeda Galaxy (upper straight line) and quasars (lower straight line) projected onto the halos of spiral galaxies. The logarithm of the galactocentric distance of each object (in kpc) is along the horizontal axis; the logarithm of the number of objects (globular clusters and quasars) per unit area of the halo ring is along the vertical axis. The distribution of globular clusters in the plane of the halos of spiral galaxies and the distribution of quasar projections onto the halos of spiral galaxies are well described by a power law with an index $\alpha \approx -2.4$.

**Fig. 6.** Radial distribution of galaxies seen through the halos of nearer galaxies (upper straight line). The density profile of QSO's projected onto the halos of elliptical galaxies (lower straight line). The logarithm of galactocentric distance of each object (2 corresponds to 100 kpc) is along the horizontal axis; the logarithm of the number of objects per unit area of the halo ring is along the vertical axis. The distribution of quasar projections onto the halos of elliptical galaxies is well described by a power law with an index $\alpha \approx -1.5$.

Let us select the close pairs from samples 1 and 2 in which the quasar is projected onto the halo of a spiral galaxy. The result is indicated by the lower straight line in Fig.5. For these quasars, $\alpha=-2.6 \pm 0.3$ in a segment up to 80 kpc. Figure 5 leads us to conclude that the distribution of globular clusters in the plane of the halos of spiral galaxies and the distribution of quasar projections onto the halos of spiral galaxies are described by a power law with a mean index $\alpha \approx -2.4$, i.e., the quasars from close quasar–galaxy pairs may follow the halo globular clusters in their radial distribution.

## 7. Globular clusters in elliptical galaxies and quasar-galaxy associatios

Let us select the quasar–galaxy pairs from samples 1 and 2 in which the quasar is projected onto an elliptical galaxy. We will construct a dependence similar to that in Fig.5. The result indicated by the lower line in Fig.6 leads us to conclude that the distribution of quasars in the halos of elliptical galaxies obeys a power law with an index $\alpha=-1.5\pm0.3$ up to 80 kpc. The quasars father than 80 kpc may be background ones unassociated with the presumed gravitational lenses of the halo.

The upper straight line in Fig.6 was constructed by analyzing the galaxy-galaxy pairs. We selected the galaxy–galaxy pairs from the SDSS catalog in a similar way as the quasar–galaxy pairs. It was necessary to determine how the radial distribution of galaxies seen through the halo of a nearer galaxy fell off. We see that the galaxies are distributed uniformly in projection onto the sky. Note that the radial distribution of galaxies seen through the halos of nearer galaxies and the distribution of quasars projected onto the halos of elliptical galaxies at distances larger than 80 kpc are identical. The distribution 263 globular clusters in the elliptical galaxy M49 within 30 kpc of its center is clearly presented in Cote et al. [15]. This galaxy contains a total of 6000 globular clusters within 100 kpc of its center. Cote et al. [14] analyzed the distribution of 278 globular clusters in the elliptical galaxy M87 within 30 kpc of its center. There are 13 500 globular clusters within 100 kpc of its center. Woodworth and Harris [33] provided the spatial distribution of globular clusters in the galaxy IC 4051. The density of globular clusters in elliptical galaxies increases toward the center. This led Blakeslee [6] to conclude that the radial density of globular clusters in the halos of elliptical galaxies A754, A1644, A2124, A2147, A2151, and A2152 is well described by a power law in form (1) with a mean $\alpha \approx -1.5$. Thus, Fig.6 suggest that the distribution of globular clusters in the plane of the halos of elliptical galaxies and the distribution of quasar projections onto the halos of elliptical galaxies are described by a power law with an index $\alpha \approx -1.5$, i.e., the quasars from close quasar–galaxy pairs follow the halo globular clusters in their radial distribution.

## 8. Conclusions and observational tests

Active galactic nuclear projected onto the halos of nearer galaxies may be seen as quasars-galaxy associations. Among them there are quasars that are close to the galaxies not only in angular separation, but also in redshift. Such quasar–galaxy pairs are called close pairs. In this paper, we developed further the hypothesis that such pairs appear, because the flux of the nucleus of the more distant galaxy passes through halo globular clusters of the nearer galaxy, resulting in magnification and splitting of the image of the source that we interpret as a quasar. To corroborate this hypothesis, we analyzed the distribution of quasars in the plane of the halos of these galaxies. The quasars from close pairs were found to follow the density profile of globular clusters in the halos of elliptical and spiral galaxies with slopes of $\alpha \approx -1.5$ and $\alpha \approx -2.4$ for elliptical and spiral galaxies, respectively. This suggests that quasars do not appear near galaxies by chance and that quasars are associated with galaxies via halo globular clusters. The quasars from close quasar–galaxy pairs can be observed to study the splitting of their images. The presumed splitting angles between the images are very small (several milliarcseconds), but they are nevertheless accessible to such telescopes as the VLTI in Chile. Another observational test consists in the following. If the quasars from pairs are actually the central sources of galaxies magnified by globular clusters in the nearer galaxy, then, since $z_{gal}/z_{qso} > 0.9$ in close pairs, the stars of the host galaxy of the quasar will be mixed with stars of the nearer lensing galaxy for an observer. This may give rise to lines in the galaxy spectra corresponding to two redshifts.

It is interesting to test a hypothesis that host galaxies of quasars close to us with z <0.3 belong to a class optical-double or interacting galaxies (see pairs in [12]). According to this hypothesis, the observer sees a quasar through halo closer galaxy, not being true host galaxy. The true host galaxy lays on the line of sight with a visible host galaxy, and probably interacts with it (as an example of such galaxy see fig.7).

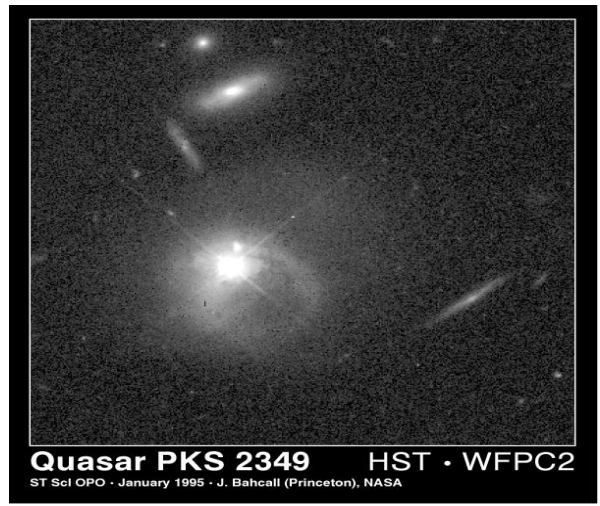

**Fig.7** The quasar PKS 2349 (the star-like object near the center) and a galaxy (surrounding fuzzy patch), but the quasar is not at the galaxy's center! In fact, the galaxy and the quasar seem to be colliding or merging. This and other recent HST observations suggest that the close quasar-galaxy pairs are interesting objects for more careful observational studies.

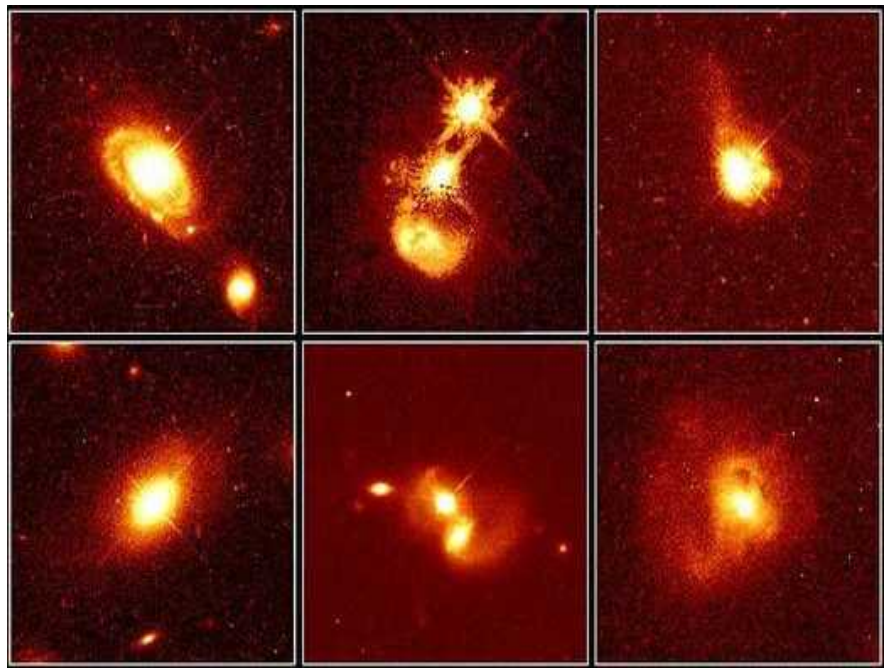

**Fig. 8** Close quasar-galaxy pairs appear in a variety cases from normal to highly disturbed systems. The column of HST images on the left represents normal galaxies; the center, colliding galaxies; and the right, peculiar galaxies: PG 0052+251, PHL 909, IRAS04505-2958, PG 1012+008, 0316-346, IRAS13218+0552 (see the text).

In Fig.8 we present several examples of close quasar-galaxy pairs (z<0.3). Top left: This image shows quasar PG 0052+251, at the core of a normal spiral galaxy. Surprisingly the host galaxy, that appear undisturbed by the strong quasar radiation, which is natural for the gravitational lensing hypothesis. Bottom left: Quasar PHL

909 lies at the core, but not in the center(!) of an apparently normal elliptical galaxy. Top center: The photo reveals evidence of a collision between two galaxies where the close quasar-galaxy pair appears.. The debris from this collision may be fueling quasar IRAS04505-2958. Bottom center: quasar PG 1012+008, merging with a bright galaxy (the object just below the quasar). providing strong evidence for an interaction between them. This may be the case of lensing of nuclear core (of the father galaxy) by the globular cluster in the halo closer galaxy. Top right: a tidal tail of dust and gas beneath quasar 0316-346, suggests that the host galaxy has interacted with a passing galaxy that is not in the image. Bottom right: evidence of merging between host galaxy of quasar IRAS13218+0552 and closer galaxy with systems of globular clusters (potential lenses) in its halo. . The elongated core in the center of the image may comprise the two nuclei of the merging galaxies. Actually we have a catalog of 130 pairs with expected similar behavior [12]

The probability to find an irregular galaxy in a pair with quasar is higher than for other types of galaxies [9]. Here we were unable to perform similar studies with irregular galaxies, because we found no close pairs among the 203 quasar–irregular galaxy pairs found in [9]. However, the following fact is of considerable interest: despite the smaller number of quasar–irregular galaxy pairs compared to the number of quasar–elliptical galaxy and quasar–spiral galaxy pairs, the relative contribution of irregular galaxies in pairs is considerably higher, i.e., among the elliptical and spiral galaxies, 1–2% of their total number are in pairs with quasars, while among the irregular galaxies, about 9% of their total number are in pairs with quasar [11].This suggest that there may be an enhanced number of compact objects with masses and radii typical of globular clusters around the irregular galaxies at distances of $50-100$ kpc. Detection of these effects would be yet another observational tests on the possibility of an association between quasars and nearby galaxies.

## References

1. J.M. Barnothy, *BAAS* **6**, 212 (1974 ).
2. Yu. V. Baryshev and Yu. L. Bukhmastova, *Pis'ma Astron. Zh.* **30**, 493(2004)[*Astron. Lett.* **30**, 444 (2004)].
3. Yu. V.Baryshev and Yu. L. Ezova, *Astron.Zh.* **74**, 497(1997) [*Astron. Rep.* **41**,436 (1997)].
4. Yu. Baryshev, A.Raikov, A.Yuschenko, in *Gravitational lenses in the universe, 31st Liege Int. Ap. Coll.*, 307(1993).
5. N. Benitez, J. L. Sanz, and E. Martinez-Gonzalez, *Mon. Not. R. Astron. Soc.* **320**,241 (2001); astro-ph/0008394.
6. J. P. Blakeslee, *Astron. J.* **118**, 1506 (1999); astro-ph/9906356.
7. J. P. Blakeslee, J. L. Tonry, and M. R. Metzger, *Astron.J.* **114**, 482, (1997).
8. P. V. Bliokh and A. A. Minakov, *Gravitational Lenses* (Naukova Dumka, Kiev,1989) [in Russian].
9. Yu. L. Bukhmastova, *Astron. Zh.* **78**, 675 (2001) [*Astron.Rep.* **45**, 581 (2001)].
10. Yu. L. Bukhmastova, *Astrofizika* **45**, 231 (2002) [*Astrophys.* **45**, 191 (2002)].
11. Yu. L. Bukhmastova, *Pis'ma Astron. Zh.* **29**, 253(2003) [*Astron. Lett.* **29**, 214(2003)].
12. Yu. L. Bukhmastova, *Pi'sma Astron. Zh.* **33**, 403(2007) [*Astron.Lett.* **33**, 355(2007)].
13. C.R.Canizares, *Nature* **291**, 620 (1981).
14. P. Cote, D. E. McLaughlin, D. A. Hanes, et al., *Astrophys.J.* **559**, 828 (2001);astro-ph/0106005.
15. P. Cote, D. E. McLaughlin, J. G. Cohen, and J. P.Blakeslee, *Astrophys. J.***591**, 850 (2003); astro-ph/0303229.
16. M. Fukugita, K. Shimasaka, and T. Ichikawa, *Publ.Astron. Soc. Pac.* **107**, 945(1995).
17. S. Galleti, L. Federici, M. Bellazzini, et al., *Astron.Astrophys.* **416**, 917 (2004); ftp://cdsarc.ustrasbg.fr/pub/cats/J/A+A/416/917; *http://cdsweb.ustrasbg. fr/viz-bin/VizieR?-source=J/A+A/416/917.*
18. W. E. Harris, *Astron. J.* **112**, 1487 (1996);www.physics.mcmaster.ca/Globular.html.
19. I. King, *Astron. J.* **67**, 471 (1962).
20. A. Klypin, S.Gottloeber, A. Kravtsov, and A. Khohlov, *Astrophys. J.* **516**, 516 (1999).
21. D.E.McLaughlin, *Astron. J.* **117**, 2398 (1999).
22. B.Menard and M. Bartelmann, *Astron. Astrophys.* **386**, 784 (2002); astro-ph/0203163.
23. B. Moor, S. Ghigna, F. Governato, et al., *Astrophys.J. Lett.* **524**, L19 (1999).
24. J.Paturel, *Astrophys. J., Suppl. Ser.* **124**, 109 (1987).
25. L.Pietronero, *Physica A* **144**, 257 (1997).
26. P.Schneider, J. Ehlers, and E. E. Falko, *Gravitational Lenses* (Springer-Verlag, New York, 1992).
27. R. Scranton, B.Menard, G. Richards, et al., *Astroph.J.* **633**, 589 (2005); astro-ph/0504510.
28. F. Sylos-Labini, M. Montuori, and L. Pietronero, *Phys. Rep.* **293**, 61 (1998).
29. J. Surdej, J.F.Claeskens et al., *AJ* 105, 2064, (1993).
30. P. Teerikorpi, *A&A* 353, 77(2000).
31. P. A. Thomas, R. L. Webster, and M. J. Drinkwater, *Astron. Astrophys.* **299**, 353 (1995); astro-ph/9408088.
32. M.P. Veron-Cetty and P. Veron, *Quasars and ActiveGalactic Nuclei ESO Sci. Rep. 18* (1998).
33. S.C. Woodworth and W. E. Harris, *Astron. J.* **119**,2699 (2000); astro-ph/0002292.
34. A.Yonehara , M. Umemura, and H. Susa, *Astron.Soc. Pac.* **55**, 1059 (2003); astro-ph/0310296.
35. A. F. Zakharov, *Gravitational Lenses and Microlenses* (Yanus-K, Moscow, 1997) [in Russian].
36. Zhu Zong-Hong, Wu Xiang-Ping, and Fang Li-Zhi, *Astrophys. J.* **490**, 31 (1997); astro-ph/9706289.